\documentclass[
 reprint,
amsmath,
amssymb,
aps,
pra,
floatfix,
]{revtex4-2}

\usepackage{graphicx}
\usepackage{dcolumn}
\usepackage{bm}
\usepackage{hyperref}
\usepackage[normalem]{ulem}
\usepackage{tikz}

\usepackage{xcolor}

\begin{document}

\preprint{APS/123-QED}

\title{Path-integral Monte Carlo estimator for the dipole polarizability of quantum plasma}

\author{Juha Tiihonen}
\email{juha.tiihonen@tuni.fi}
\affiliation{%
Faculty of Engineering and Natural Sciences, Tampere University, Tampere, Finland
}%

\author{David Trejo-Garcia}
\affiliation{%
Faculty of Engineering and Natural Sciences, Tampere University, Tampere, Finland
}%

\author{Tapio T. Rantala}
\affiliation{%
Faculty of Engineering and Natural Sciences, Tampere University, Tampere, Finland
}%

\author{Marco Ornigotti}
\affiliation{%
Faculty of Engineering and Natural Sciences, Tampere University, Tampere, Finland
}%

\date{\today}

\begin{abstract}
We present a path-integral Monte Carlo estimator for calculating the dipole polarizability of interacting Coulomb plasma in the long-wavelength limit, i.e., the optical region. We present comprehensive details and method validation studies for our approach based on both collective and one-particle dipole autocorrelation functions in the imaginary time. The simulation of thermal equilibrium in imaginary time has exact Coulomb interactions and Boltzmann quantum statistics. For reference, we use analytically continued Drude model as the long-wavelength limit of the Lindhard response. Our collective response shows perfect match to the analytical reference. The one-particle response is used in systematic studies of physical and numerical parameters, and to discuss the phenomenological Drude scattering model. 
\end{abstract}

\maketitle

\section{Introduction}
\label{sec:intro}

The Drude model, introduced at the beginning of the last century by Paul Drude as an attempt at a comprehensive electron theory of metals \cite{drude1,drude2}, remains nowadays ubiquitous in describing the optical response of metals, as well as optical materials with metal-like behaviours, such as epsilon-near-zero (ENZ) media \cite{enz1}. The Drude parameters have been fitted to experiments \cite{Johnson1972} and hydrodynamic models of various kinds typically recover the Drude model or something close to it \cite{Maack2018}. One of the advantages of the Drude model, beside its simplicity, resides in its high effectiveness in describing the optical response in a broad range of frequencies, from the near IR to the UV regime. This allows to characterize the behaviour of different photonic materials, such as metals \cite{ashcroft, Singwi1968}, plasmonic circuits \cite{maier}, ENZ materials \cite{enz2}, as well as more complicated photonics platforms, such as metasurfaces \cite{Schulz2024}, with an effective Drude model. Yet, the Drude model remains phenomenological and unable to predict all subtleties of quantum mechanics from first principles.

In spirit, the Drude model describes the dynamics of non-interacting free electrons, which occasionally scatter from ions. When the scattering is modest, this model becomes the homogeneous electron gas (HEG), or jellium. Perhaps the most established framework for treating the dielectric response of HEG is the Lindhard theory, which builds upon the Lindhard function $\chi_0(\mathbf{q}, \omega)$ \cite{Giuliani2008}, where $\mathbf{q}$ and $\omega$ are the respective wave number and frequency of an incoming perturbation. While somewhat tedious to evaluate numerically, $\chi_0$ recovers many physical insights in the right limits. In particular, the Drude model is recovered by taking $\chi_0$ to the limits of long wavelength and low scattering \cite{Singwi1968, ashcroft}. 

The Lindhard function cannot also describe the full many-body interactions of HEG, unless the Lindhard response is expressed more generally, like \cite{kugler1975theory, Giuliani2008}
\begin{align}
    \label{eq:chi_q}
    \chi(\mathbf{q}, \omega) &= -v_q \chi_{nn}(\mathbf{q}, \omega) \\
    \label{eq:rpa}
    \chi_{nn}(\mathbf{q}, \omega) &= \frac{\chi_0(\mathbf{q}, \omega)}{1 + v_q G(\mathbf{q}, \omega) \chi_0(\mathbf{q}, \omega)},
\end{align}
where $\chi$ denotes the complex dielectric susceptibility, $\chi_{nn}$ is the screened density--density susceptibility, $v_q = e^2/|\mathbf{q}|^2 \varepsilon_0$ is the reciprocal Coulomb potential (in SI units) and $G(\mathbf{q}, \omega)$ is the so-called local field correction (LFC) treating many-body effects due to exchange and correlation. When $G=0$, Eq.~\eqref{eq:rpa} reduces to $\chi_0(\mathbf{q}, \omega)$ and amounts to the random phase approximation (RPA). For decades, numerous works (e.g. Refs.~\cite{kugler1975theory, Moroni1995, Groth2019, Hamann2020, Dornheim2023a}) have discussed and featured either static $G(\mathbf{q}, \omega=0)$ or dynamic $G(\mathbf{q}, \omega)$ in pursuit of the non-trivial response within the so-called regime of warm, dense matter (WDM). 

Recently, significant progress has been made to calculate the density response of the WDM with the path-integral Monte Carlo method (PIMC) \cite{Dornheim2018,Groth2019,Dornheim2020a,Dornheim2021,Dornheim2021a,Dornheim2023a,Chuna2025,Dornheim2022}. The PIMC approach can treat accurate quantum many-body interactions at finite temperatures, and it is in principle only subject to numerical challenges due to the Fermion sign problem (FSP) \cite{Loh1990, Ceperley1995}. Within this framework, moreover, the dielectric response is directly connected to the dynamic structure factor through the intermediate scattering function \cite{Giuliani2008}. The latter can be straightforwardly estimated from the PIMC walker as the density-density correlation function in the imaginary time. This function can be then analytically continued in the complex plane to recover the full dynamic response of the system. The inverse transformation back to the real time, however, faces serious challenges with numerical noise, prevalent in Monte Carlo simulation, and thus requires special methods \cite{Jarrell1996, Bergeron2016, Chuna2025}. Within this framework, Dornheim et al. have comprehensively studied the first \cite{Dornheim2023a} and higher order \cite{Dornheim2020a, Dornheim2021} response properties of the HEG. An apparent limitation of their approach is that the allowed values for the wavenumber are limited by $|\mathbf{q}| = 2 \pi n / L$, where $n \geq 1$ and $L$ is the (cubic) simulation cell size. Therefore, the typical optical wavelengths with $q \ll L^{-1}$, thus requiring large $L$, are hard to accomplish because of the algorithmic scaling and performance loss attributed to the high particle count $N$. Indeed, the literature focuses on characterizing inelastic neutron scattering or X-ray Thompson scattering spectra, where the momentum transfers are significant \cite{Dornheim2023a} compared to that of photons.

Here, we explore a complementary PIMC approach to regain focus on the long, optical wavelengths. We, in fact, demonstrate computation of the conventional optical response based on the complex dipole polarizability, or the dipole-dipole correlation function. This approach, which evades the difficulty of low $\mathbf{q}$, has been studied before in finite systems \cite{Gallicchio1994, Gallicchio1996, Tiihonen2018, Tiihonen2019} but not in periodic ones. The long-wavelength approximation, where we treat a constant field in space, is well justified for optical region: For example, the box size $L$ required for $N=16$ particles to simulate the metallic density of gold is over $300$ times smaller than an ultraviolet wavelength $\lambda = 200$ nm. Treating this naïvely in the Fourier space would require $N \sim 10^{8}$ particles in a cubic box.

However, it is not obvious \textit{a priori} how a PIMC estimator for the dipole polarizability must be laid out for periodic systems or how it responds to finite-size effects or converge parameters, such as the finite time-step. For these reasons, this work aims at establishing these matters in a controllable setting. To do so, we first discuss analytical correspondence between the Drude model and the PIMC polarizability, evaluated in the imaginary time. We then present ways to estimate the dipole polarizabilities of individual and collective particles from the periodic simulation of a finite volume. Then, we survey how the results based on individual and collective properties respond to physical and methodological parameters. The collective picture matches the target, while the one-particle observable is found with clear many-body signatures, strongly influenced by the physical environment. We use this information to discuss an effective screening picture through a phenomenological Drude damping approximation.

This work is organised as follows: In Sec.~\ref{sec:theory}, we lay out the theoretical context of the estimator and the PIMC simulation. In Sec.~\ref{sec:methods} we overview the computational details. In Sec.~\ref{sec:results} we discuss the numerical results and in Sec.~\ref{sec:summary} we summarize the work and future prospects of this methodology.

\section{Theory}
\label{sec:theory}

In this section, we first overview the linear optical response of quantum particles in the long-wavelength limit. We then discuss estimation of the dipole--dipole correlation function from imaginary-time PIMC simulation. Finally, we consider the Drude model as a non-interacting reference theory and its analytic continuation into the imaginary domain. The results of this section are presented in SI units, but the reader should keep in mind that later results will be given in in Hartree atomic units, i.e.,  $\hbar = 4 \pi \varepsilon_0 = m_e = 1$, for convenience.

\subsection{Electric field response}
\label{sec:ef_response}

Let us consider a quantum mechanical plasma that is perturbed by a weak external electric field $\mathbf{E}(\mathbf{r'}, \omega)$ such that the Hamiltonian becomes
\begin{equation}
    H \rightarrow H_0 + H_E(\omega),
\end{equation}
where $H_0$ is the non-perturbed Hamiltonian and $H_E(\omega)$ is a field-induced energy shift, integrated over volume $V$, is
\begin{align}
    \label{eq:HE}
    H_E(\omega) &= \int_V d\mathbf{r} U_E(\mathbf{r}) \\
    \label{eq:U}
    U_E(\mathbf{r}) &= \mathbf{P}(\mathbf{r}, \omega) \cdot \mathbf{E}(\mathbf{r}, \omega),
\end{align}
where the induced electric polarization is prescribed to be linear with the (weak) field:
\begin{equation}
    \mathbf{P}(\mathbf{r}, \omega) \cong \varepsilon_0 \int_V d\mathbf{r}' \chi(\mathbf{r}, \mathbf{r}', \omega) \mathbf{E}(\mathbf{r'}, \omega),
\end{equation}
where $\varepsilon_0$ the vacuum permittivity and $\chi(\mathbf{r}, \mathbf{r}', \omega)$ is the dielectric susceptibility tensor. 

By assuming translation invariance of plasma and regularity of $\mathbf{E}(\mathbf{r})$, we may write $\chi(\mathbf{r}, \mathbf{r}', \omega)=\chi(\mathbf{r}-\mathbf{r}', \omega)$ and consider its Fourier transform $\chi(\mathbf{q}, \omega)$ like in Eq.~\eqref{eq:chi_q}. On the other hand, we can make a long-wavelength approximation in the locality $L \sim V^{-3}$ such that $L \sim |\mathbf{r}-\mathbf{r'}| \ll \lambda$, where $\lambda = 2\pi/|\mathbf{q}|$ is the wavelength of the locally constant field $\mathbf{E}(\mathbf{r} - \mathbf{r}', \omega) \cong \mathbf{E}(\omega)$. Thus, we can re-express Eq.~\eqref{eq:U} as a uniform potential energy density with
\begin{align}
    \delta U_E(\omega) &\cong \frac{1}{V} \int_V d(\mathbf{r}-\mathbf{r'}) \mathbf{P}(\mathbf{r} - \mathbf{r'}, \omega) \cdot \mathbf{E}(\omega)\\
    \label{eq:deltaU1}
    &= \varepsilon_0 |\mathbf{E}(\omega)|^2 \chi(\omega)
\end{align}
where $\chi(\omega) \equiv V^{-1} \int_V d(\mathbf{r}-\mathbf{r'}) \chi(\mathbf{r} - \mathbf{r'}, \omega)$ is the bulk dielectric susceptibility, describing linear optical dispersion in the long-wavelength limit through the canonical relations \cite{Jackson1975}
\begin{align}
    \varepsilon(\omega) &= 1 + \chi(\omega) \\
    \mathbf{D}(\omega) &= \varepsilon_0 \varepsilon(\omega) \mathbf{E}(\omega),
\end{align}
where $\varepsilon$ is the complex dielectric function and $\mathbf{D}(\omega)$ is the total electric displacement field.

\subsection{Quantum response functions}
\label{sec:quantum_response}

Let us consider charged plasma comprising of $N$ identical quantum particles with charge $-e$ in a cubic box of volume $V=L^3$. For convenience, and without loss of generality, we can reduce the rank-2 susceptibility tensor $\chi(\omega)$ to its spherically symmetric diagonal, denoted $z$, and consider the response of this (homogeneous, isotropic) system to a long-wavelength electric field $E_z(\omega)$ applied along the direction $z$. From the zero-field limit of Eq.~\eqref{eq:deltaU1}, we can derive
\begin{equation}
    \varepsilon_0 \chi(\omega) = \left. \frac{\partial^2}{\partial E_z^2} \frac{\delta U_E(\omega)}{V} \right|_{E_z \rightarrow 0},
\end{equation}
which leads, using $U_E(\mathbf{r}, \omega) = \mu(\mathbf{r}) \cdot \mathbf{E}(\omega)$ and perturbation theory, to the Kubo formula \cite{Kubo1957}
\begin{align}
    \label{eq:chi_w}
    \chi(\omega) &= -\frac{i n}{\hbar \varepsilon_0} \int_0^\infty \mathrm{d}t e^{i\omega t} 
    \langle [\mu_z(t), \mu_z(0)] \rangle / N,
\end{align}
where $n=N/V$ is the number density, the angle brackets evaluate a quantum statistical average over the volume $V$, and square brackets denote a commutator. The dipole moment operator in $z$ direction is generally
\begin{equation}
    \label{eq:dipole_moment}
    \mu_z(\mathbf{r}, t) = \sum_{i=1}^N q_i z_i \delta(\mathbf{r} - \mathbf{r_i}(t)),
\end{equation}
where $q_i$, $\mathbf{r \in} V$ and $\mathbf{r_i}(t)$ are the respective charge and position of the $i$th particle. If $|q_i|=e$ for all $i$, Eq.~\eqref{eq:chi_w} can be expressed with the Fourier-space density--density susceptibility $\chi_{nn}(\omega)$ like in Eq.~\eqref{eq:chi_q} or the dynamic structure factor \cite{Giuliani2008}, as is typically done to treat shorter wavelengths.

Let us denote the dipole--dipole correlation function with
\begin{equation}
    G(t) \equiv \frac{i}{\hbar} \langle [\mu(t), \mu(0)] \rangle/N
\end{equation}
such that the retarded (causal) correlation function is
\begin{equation}
    G^R(\omega) = \int_{-\infty}^{\infty} \mathrm{d}t e^{i\omega t} \Theta(t) G(t) = -\frac{\varepsilon_0}{n} \chi(\omega),
\end{equation}
whose spectral density function is \cite{Jarrell1996}
\begin{equation}
    \label{eq:spectrum}
    A(\omega) \equiv -2 \mathrm{Im}[G^R(\omega)].
\end{equation}
The spectral density allows analytic continuation of $G^R$ to imaginary time $-it \rightarrow \tau$ through \cite{Jarrell1996}
\begin{equation}
    \label{eq:Aw_to_Gtau}
    \mathcal{G}(\tau) = \int_{-\infty}^{\infty} \frac{d \omega}{2 \pi} \frac{\mathrm{e}^{-\tau\omega}}{1 - \mathrm{e}^{-\beta \omega}} A(\omega),
\end{equation}
where $\beta = 1 / k_B T$ is the inverse temperature. The imaginary-time correlation function $\mathcal{G}(\tau)$ has bosonic symmetry with $\mathcal{G}(\tau) = \mathcal{G}(\tau + n\beta)$ for any integer $n$, and similar to its counterpart $G^R(t)$, is evaluated with 
\begin{equation}
    \label{eq:Gtau}
    \mathcal{G}(\tau) = \frac{1}{\hbar} \langle \mathcal{T} \mu_z(0) \mu_z(\tau) \rangle/N,
\end{equation}
where $\mathcal{T}$ is the time-ordering operator. The dipole moment operators $\mu_z(\tau)$ are also assumed normal ordered, meaning $\langle \mu_z(\tau) \rangle = 0$. The Fourier transform of $\mathcal{G}(\tau)$ is the Matsubara series $\mathcal{G}(i \omega_n)$, where $\omega_n = 2 \pi n / \hbar \beta$ for all integers, which can also be obtained with \cite{Jarrell1996}
\begin{equation}
    \label{eq:Aw_to_Giw}
    \mathcal{G}(i \omega_n) = -\int_{-\infty}^{\infty} \frac{d \omega}{2 \pi} \frac{1}{i\omega_n - \omega} A(\omega).
\end{equation}

While the imaginary-domain functions $\mathcal{G}(\tau)$ and $\mathcal{G}(i\omega_n)$ contain the same information as $G^R(\omega)$, the conversion from imaginary to real domain by inverting Eqs.~\eqref{eq:Aw_to_Gtau} or \eqref{eq:Aw_to_Giw} is a known ill-posed problem, which is accentuated by the unavoidable error bars present in quantum Monte Carlo methods \cite{Jarrell1996}. The use of existing numerical methods of analytic continuation, such as Maximum Entropy \cite{Bergeron2016, Gallicchio1994}, will not be pursued here.

\begin{figure*}[ht]
   \centering
   \begin{tikzpicture}
   \node (wind) at (0, 0) {\includegraphics[height=6.0cm]{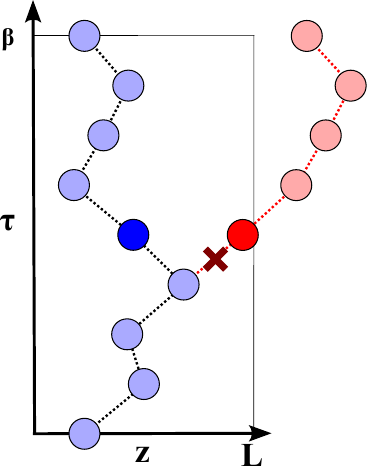}};
   \node at (wind.north west) {(a)};
   \node (unfold) at (4.5, 0) {\includegraphics[height=6.0cm]{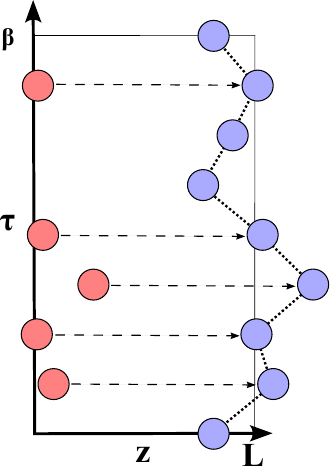}};
   \node at (unfold.north west) {(b)};
   \node (dm1) at (8.7, 0) {\includegraphics[height=6.0cm]{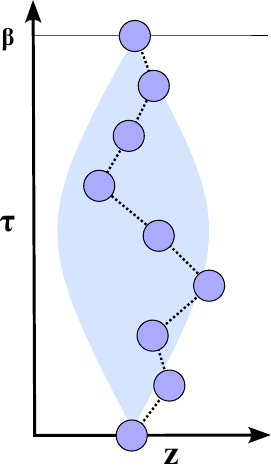}};
   \node at (dm1.north west) {(c)};
   \node (collective) at (12.7, 1.5) {\includegraphics[height=2.0cm]{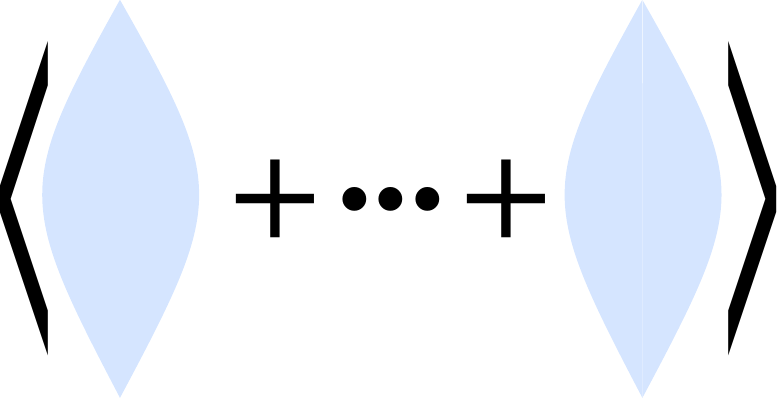}};
   \node at (collective.north west) {(d)};
   \node (1particle) at (12.7, -1.5) {\includegraphics[height=2.0cm]{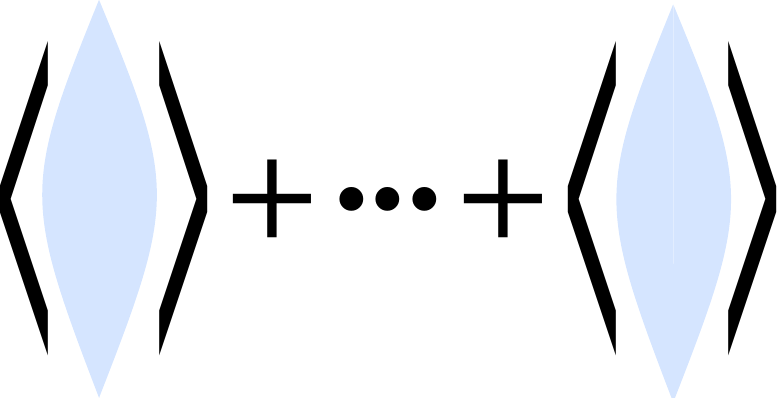}};
   \node at (1particle.north west) {(e)};
   \end{tikzpicture}
   \caption{
   Schematic illustrations of the periodic dipole moment estimator and sampling constraints. The blue and red circles represent Monte Carlo walkers of quantum particles, i.e., positions advancing in discrete steps through imaginary time and forming closed loops $z(0) = z(\beta)$ modulo $L$, where $z$ denotes the real coordinate in one dimension. (a) Winding restriction. MC sampling may cause the path to wind across box boundary (branch of red dots), when the thermal wavelength of a multilevel bisection move is a significant fraction of $L$. For distinguishable particles, this is clearly an artifact of the finite cell and also detrimental to the estimation of $\mathcal{\tilde G}(\tau)$, and thus, such moves are rejected. (b) Unfolding of the path. Since the estimator does not treat periodicity with a Fourier phase, paths crossing the simulation box boundary to the nearest image must be unfolded, i.e., moved within $L/2$ of the earlier bead, before measurement. (c) Dipole measurement. The autocorrelation of induced dipole moment is easily measured from the mean squared deviation of the instantaneous dipole moment at $\mu(\tau)$ relative to $\mu(0)$ according to Eq.\eqref{eq:relative_estimator}. (d) The collective response is obtained when the instantaneous fluctuations of all particles are summed up \textit{before} correlating $\mu(0)$ with $\mu(\tau)$. (e) The one-particle response is obtained when the instantaneous fluctuations of all particles are summed up \textit{after} correlating $\mu_i(0)$ with $\mu_i(\tau)$.}
   \label{fig:paths}
\end{figure*}

\subsection{PIMC estimator}
\label{sec:pimc}

The quantum statistical averages of the dipole--dipole correlation function at finite temperatures can be obtained with the canonical PIMC method \cite{Ceperley1995}. The method draws samples from the distribution of the thermal density matrix:
\begin{equation}
    \label{eq:density_matrix}
    \rho(R, R; \beta) = Z^{-1} \int dR \langle R | \hat{\rho}(\beta) | R \rangle,
\end{equation}
where $R$ denotes a set of variable particle coordinates, $\hat{\rho}(\beta) = e^{-\beta H_0}$ is the density operator at inverse temperature $\beta$, $H_0$ is the Hamiltonian and $Z = \langle R | \hat{\rho} | R \rangle$ is the thermal partition function. Equation~\eqref{eq:density_matrix} overlooks summations arising from spin degrees of freedom and particle permutations required to treat proper quantum statistics of bosons or fermions; instead, the simulation treats so-called Boltzmannons, which comprise a Coulomb interacting gas of distinguishable quantum particles \cite{Ceperley1986}. The Boltzmannons evade the FSP but do not correspond to physical electrons, alas, except at high temperatures and low densities \cite{Ceperley1995} (or strong coupling \cite{Dornheim2016}), where particle repulsion outweighs exchange and the quantum statistics merge. As elaborated in the SI, some of our results indeed start to deviate from the Fermion values.

The interacting Hamiltonian for $N$ distinguishable particles is given by
\begin{align}
    H_0 = K + U = \hbar^2 \lambda_m \sum_i \nabla_i^2 + \sum_{i < j} \frac{q_i q_j}{4 \pi \varepsilon_0|\mathbf{r}_i - \mathbf{r}_j|}
\end{align}
where $\lambda_m = 1/2 m_e$ and $K$ and $U$ correspond, respectively, to the summation over kinetic and potential operators, the last ones comprising exact Coulomb interaction between pairs of particles. Since $K$ and $U$ do not commute, the density operator must be expanded with the semi-group property
\begin{equation}
    \label{eq:trotter_expansion}
    e^{-\beta(K + U)} = \lim_{M \rightarrow \infty} \left[ e^{-\Delta \tau K} e^{-\Delta \tau U} \right]^M,
\end{equation}
where $\Delta \tau = \hbar \beta / M$ is a finite time-step and $M$ is the Trotter number. Using Eq.~\eqref{eq:trotter_expansion} and the product property \cite{Ceperley1995}, the exact Coulomb action between pairs of species can be treated for finite time-steps \cite{Kylaenpaeae2011}. In this so-called pair approximation, higher-body interactions retain a modest time-step error proportional to $(\Delta \tau)^3$ \cite{Brown2013, Kylaenpaeae2011}. Therefore, it is justified to choose finite $M$, making the PIMC integrand a discrete walker of $M$ times $R$ coordinates, which forms a closed loop, an imaginary-time trajectory: 
\begin{align}
    \label{eq:discrete_path}
    \rho(R, R; \beta) = Z^{-1} \int \prod_{j=1}^{M} dR_j \langle R_{j-1} | \hat{\rho}(\Delta \tau) | R_j \rangle,
\end{align}
where $R_j$ contains the real-space coordinates for $N$ particles at $j$th time-slice, and the loop closes with $R_M = R_0$. The trajectory is efficiently sampled as a Markov chain using the Metropolis multilevel bisection algorithm, as described in \cite{Ceperley1986,Ceperley1995,Kylaenpaeae2011}.

Thermal expectation values of diagonal properties, such as potential energy $V$, is can be calculated from
\begin{align}
    \langle U \rangle &= Z^{-1} \int dR \langle R | U \hat{\rho}(\beta) | R \rangle \\
    &\approx Z^{-1} \int \left[ \prod_{j=1}^{M} dR_j \rho \left(R_{j-1}, R_j; \Delta \tau \right) \right] U(R_0) \\
    &= \frac{1}{M} \sum_{j=1}^M \langle U(R_j) \rangle,
\end{align}
where the last line employs symmetry properties of the trace, enabling to calculate the mean over all time-slices of any trajectory. Because of the kinetic operator, the total energy $E$ is more complicated; here we use the thermal estimator \cite{Ceperley1995}, whose numerically exact pair energy $-\partial  \rho/\partial \beta$ can be precomputed \cite{Kylaenpaeae2011}.

The calculation of time-ordered observables such as Eq.~\eqref{eq:Gtau} is equally straightforward. The mean per-particle correlation function can evaluated at discrete multiples of $\Delta \tau$ by \cite{Tiihonen2019}
\begin{align}
    \label{eq:Gtau_pimc}
    \langle \mathcal{G}(\tau) \rangle = \frac{1}{\hbar ZN} \prod_{j=1}^{M} \int dR_j \rho(R_{j-1}, R_j; \Delta \tau) \mu_z(R_0) \mu_z(R_{t}),
\end{align}
where the integer time-difference index is $t = j + \tau / \Delta \tau$, modulo $M$. A snapshot measurement of the correlation function from the trajectory is visualized in Fig.~\ref{fig:paths} (left). The expression Eq.~\eqref{eq:Gtau_pimc} can be easily symmetrized, optimized for evaluation and generalized for different combinations of measurement operators, as detailed in \cite{Tiihonen2018, Tiihonen2019}.

The imaginary-time dipole moment operating on a time-slice is
\begin{align}
    \label{eq:dm_collective}
    \mu(R_t) &= \sum_{i=1}^N \mu_i(R_t), \\
    \label{eq:dm_1particle}
    \mu_i(R_t) &= q_i \left(z_t - z_0\right)
\end{align}
where $\mu_i$ symbolizes the dipole moment of particle $i$ and $z_0$ is a formal reference point for $\mu$. A difficulty arises for free plasmas, for which $\sum_i q_i \neq 0$ and there is no extrinsic reference point because of the translation invariance. One might be tempted to use an intrinsic reference point, such as the path-centroid $z_0 \equiv M^{-1} \sum_t z_t$, but that leads to $\langle \mathcal{G}(i \omega_n = 0) \rangle= 0$. Thus, like in Ref.~\cite{Gallicchio1994} we instead extract the induced dipole autocorrelation from
\begin{equation}
    \label{eq:relative_mu2}
    \mu(0)\mu(\tau)  = 2 \mu(0)^2 - (\mu(0) - \mu(\tau))^2/2,
\end{equation}
where the subtraction of $\mu(0)^2 = \mu(\tau)^2$ eliminates treatment of the static limit.
Combining Eqs.~\eqref{eq:relative_mu2} and \eqref{eq:dm_collective} yields the collective dipole autocorrelation function (see Fig.~\ref{fig:paths} (d))
\begin{align}
    \label{eq:relative_estimator}
    \mathcal{\tilde G}(\tau) &\equiv \mathcal{G}(\tau) - \mathcal{G}(0) \\
    &= -\frac{1}{2\hbar} \langle (\mu(0) - \mu(\tau))^2 \rangle/N,
\end{align}
which arises from mean squared fluctuations of the instantaneous dipole moment along the imaginary-time path (the green path in Fig.~\ref{fig:paths} (left)). This is not such a huge compromise, as the relative correlation functions in real domain follow
\begin{align}
    \tilde G^R(t) &\equiv G^R(t) - G^R(0) \\
    \tilde G^R(\omega) &= \int dt \mathrm{e}^{i \omega t} \tilde G^R(t) = -\varepsilon_0\chi(\omega \neq 0)/n,
\end{align}
so we can, in principle, treat all values of $\chi(\omega)$ except the static value.

Alternatively, combining Eqs.~\eqref{eq:relative_mu2} and\eqref{eq:dm_1particle} yields the one-particle response (see Fig.~\ref{fig:paths} (e)))
\begin{align}
    \label{eq:1particle_estimator}
    \langle \mathcal{G}_1(\tau) \rangle =& \frac{1}{N\hbar} \sum_i \langle \mu_i(0) \mu_i(\tau) \rangle \\
    \label{eq:1particle_relative}
    \langle \mathcal{\tilde G}_1(\tau) \rangle =& -\frac{1}{2 N \hbar} \sum_i \langle (\mu_i(0) - \mu_i(\tau))^2 \rangle.
\end{align}
The difference between Eqs.~\eqref{eq:relative_estimator} and \eqref{eq:1particle_relative} is
\begin{equation}
    \label{eq:1particle_diff}
    \langle \mathcal{\tilde G}(\tau) \rangle - \langle \mathcal{\tilde G_1}(\tau) \rangle = \sum_{i\neq j} \langle (\mu_i(\tau) - \mu_i(0)) (\mu_j(\tau) - \mu_j(0) \rangle,
\end{equation}
that is, the mean correlation between the dipole response of one particle and its environment. Eqs.~\eqref{eq:1particle_estimator} and \eqref{eq:1particle_diff} correspond to respective integrals over the self and distinct van Hove functions. Both observables are subject to quantum many-body effects, although the many-body signals of the collective response are completely cancelled out by the perfect screening property of ideal plasmas \cite{Giuliani2008}, which can be used to assert numerical validity of simulations. On the other hand, the one-particle response lends itself for gauging both the numerical and physical natures of the underlying many-body effects.

\subsection{Periodic boundary conditions}
\label{sec:pbc}

The simulation of charged plasma requires periodic boundary conditions (PBC). Here we use a cubic box of size $L$ (in practice like in Sec.~\ref{sec:ef_response} although semantically different), whose size is determined by the
\begin{equation}
    L = r_s \left( \frac{4 \pi N}{3}\right)^{1/3},
\end{equation}
where $r_s$ is the Wigner-Seitz radius. The treatment of periodic Coulomb images is treated using the ordinary Ewald summation \cite{Fraser1996, Natoli1995} (see SI for details). Evaluation of the relative dipole autocorrelation function (Eqs.~\eqref{eq:Gtau_pimc} and \eqref{eq:relative_estimator}), is done on walkers whose all particle paths are unfolded into continuous trajectories near boundaries, as illustrated in Fig.~\ref{fig:paths} (middle). Since the simulated particles may not exchange paths (permute), non-zero winding of a path through the cell is an artifact of the finite size. Therefore, we reject all moves leading to non-zero winding (see Fig.~\ref{fig:paths} (right)). Beside this constraint, the sampling is done using the multilevel bisection method \cite{Ceperley1995}. The probability of winding increases when the thermal de Broglie wavelength of the bisection move approaches the cell dimension, i.e.
\begin{equation}
    \sqrt{2^l \lambda_m \Delta \tau} \sim L,
    \label{eq:multilevel}
\end{equation}
where $l$ is the bisection multilevel. We constrain $l$ so that the l.h.s. of Eq.~\eqref{eq:multilevel} is lower than $L/8$. This is a simplification to avoid using periodic free-particle density matrix \cite{Ceperley1995, Kylaenpaeae2011}, which would not inhibit the winding, anyway. As a result, fewer than one winding move per $10^3$ trial moves needs to be rejected in any density. A modest sampling bias is mitigated along other finite-size effects when $L$ naturally grows with $N^{1/3}$.

\subsection{Analytic continuation of the Drude model}

The ubiquitous model for considering the optical response free-carrier species like electrons in metals is the Drude model, whose dielectric susceptibility is
\begin{equation}
\label{eq:varepsilon_D}
    \varepsilon^D(\omega) = 1 - \frac{\omega_p^2}{\omega^2 + i\omega \Gamma},
\end{equation}
and thus, the susceptibility is
\begin{align}
    \label{eq:chi_D}
    \chi^{D}(\omega) = -\frac{\omega_p^2}{\omega^2 + i\omega \Gamma},
\end{align}
where $\omega_p^2 = \tfrac{n e^2}{m \varepsilon_0} = \tfrac{e^2}{4 \pi \varepsilon_0 m} \tfrac{3}{r_s^3}$ is the plasma frequency and $\Gamma > 0$ is a positive damping coefficient. As is well known \cite{ashcroft} and iterated in Appendix A, the Drude model is recovered from the Lindhard function $\chi_0(\mathbf{q}, \omega)$ in the limits of small $\Gamma$ and long wavelength as $q \rightarrow 0$. Thus, the Drude model is used in this work as a reference model that derives from quantum mechanics and contains screening on mean-field level \cite{Giuliani2008} but neglects many-body effects from exchange and correlation. Differences between the PIMC data and Drude model are therefore manifestations of the many-body effects.

Based on Eqs.~\eqref{eq:chi_w}, \eqref{eq:chi_D} and \eqref{eq:spectrum}, the Drude spectral density is given by
\begin{align}
    \label{eq:drude_spectrum}
    A^{D}(\omega) &= \frac{2 \omega_p^2 \varepsilon_0}{n} \mathrm{Im}[\chi^D(\omega)] \\
    &= \frac{2 e^2}{m} \frac{1}{\omega (\Gamma + \omega^2 / \Gamma)}.
\end{align}
From Eq.~\eqref{eq:drude_spectrum} it is clear that the Drude spectrum does not depend on the physical density, because the oscillators are not interacting; rather, the density only shows in the transformation to the macroscopic susceptibility with Eq.\eqref{eq:chi_w}. The spectrum also diverges at $\omega = 0$ complicating its analytic continuation into $\mathcal{G}$. However, the relative correlation function $\mathcal{\tilde G}$ can be obtained, because the subtraction of $G^R(\tau=0)$ controls the divergence at $\omega=0$. So, based on Eqs.\eqref{eq:drude_spectrum}, \eqref{eq:Aw_to_Gtau} and \eqref{eq:Aw_to_Giw} we can perform numerical analytic continuation with
\begin{align}
    \mathcal{\tilde G}^D(\tau) &= -\int_{-\infty}^\infty \frac{d\omega}{2 \pi} \frac{1 - \mathrm{e}^{-\tau \omega}}{1 - \mathrm{e}^{-\beta \omega}} A^D(\omega).
\end{align}
The limit of low damping can be associated with the mean squared dipole fluctuation of a 1-dimensional free-particle density matrix, yielding a closed-form expression
\begin{align}
    \label{eq:analytical_GtauD}
    \lim_{\Gamma \rightarrow 0} \mathcal{\tilde G}^D(\tau) &= -\frac{e^2}{2 \hbar Z} \int z^2  \mathrm{exp}\left(-\tfrac{z^2}{4 \lambda_m} (\tfrac{1}{\tau}+\tfrac{1}{\beta - \tau}) \right) \mathrm{d}z \\
    &= \frac{e^2}{\hbar \beta} \left(\tau^2 - \beta \tau \right)
\end{align}
to be used as a reference.

For the Matsubara data we get an analytic expression for the non-zero Matsubara frequencies:
\begin{align}
    \mathcal{\tilde G}^D(i \omega_n \neq 0) &= \mathcal{G}^D(i \omega_n \neq 0) \\
    \label{eq:analytical_GiwD}
    &= \frac{2 \pi e^2}{m} \left[ \frac{1}{\omega_n ^2 / \Gamma +\omega_n }  \right],
\end{align}
However, we can formally define the static limit based on the zero-frequency Fourier transform
\begin{equation}
    \label{eq:static_matsubara}
    \mathcal{\tilde G}(i \omega_n=0) \equiv -\int_0^\beta d \tau \mathcal{\tilde G}(\tau),
\end{equation}
where the sign has been changed for convenience, to keep all values of $\mathcal{\tilde G}(i \omega_n) > 0$ in figures. As discussed before, $\mathcal{\tilde G}(i \omega_n=0)$ is not physical but since it describes the first-order scaling of the correlation function, we will use it later as a measure of quantity.

\section{Computational details}
\label{sec:methods}

The PIMC simulation software implements the canonical Metropolis Monte Carlo algorithm for a PIMC walker in Fortran90, featuring matrix squaring of the exact Coulomb pair action, parallel sampling and data binning of a number of Markovian walkers. Here, 40 CPU cores are used for the parallel sampling. Raw data after sample binning is stored in hierarchical data format (HDF5) and data processing chains operate on standard numerical Python libraries. The integral transformations, such as Eq.~\eqref{eq:Aw_to_Gtau} and Eq.~\eqref{eq:Aw_to_Giw}, are carried out using numerical quadrature. Fourier transforms from $\mathcal{\tilde G}(\tau) \rightarrow \mathcal{G}(i \omega_n)$ are carried on regular grids of finite spacing using fast Fourier transform (FFT); however, FFT biases the result toward higher frequencies, and thus, the original data is interpolated 16-fold while truncating the size of the result array to mitigate the bias \cite{Bergeron2016, Tiihonen2018}. The simulation campaigns are operated with Nexus workflows \cite{Krogel2016} on various high-performance computing facilities listed in Sec.~\ref{sec:acknowledgements}. Central implementation details, including code snapshots and workflow scripts for step-by-step reproduction of the simulations, postprocessing and individual figures are supplied in a separate data repository \cite{Repository}.

Results of the PIMC simulation, \textit{i.e.} energies and correlation functions, have finite statistical uncertainties due to finite sampling. The uncertainties are estimated based on 2$\sigma$ standard error of the mean (2SEM), including sample autocorrelation time $\kappa$, as described in Refs.~\cite{Kylaenpaeae2011, Tiihonen2019}. The effective means and $\kappa$ per each observable, including $\mathcal{\tilde G}(\tau)$ for each individual $\tau$, are analysed from the sequential distribution of sample block averages. The block averages are means over numerous measurements, aggregated in parallel from independent MC walkers and also by binning subsequent measurements from each walker. Within each walker, a number of MC moves is performed between each measurement to decrease sample autocorrelation. Statistical uncertainty of the derived quantity $\mathcal{G}(i \omega_n)$ is obtained with effective sample covariance by measuring 2SEM for each $\omega_n$ after Fourier transforming each block average of $\mathcal{\tilde G}(\tau)$.

\section{Results}
\label{sec:results}

The PIMC simulation, as laid out in earlier section, is applied in quantum plasmas at various finite temperatures and densities. With $m=m_e=1$ and $q^2=e^2=1$, the simulated particles resemble free electrons except with the Boltzmann statistics, as discussed in Sec.~\ref{sec:pimc}. Thus, we use here conventional units of the HEG and express densities with $r_s$ and finite temperatures with $\Theta=T/T_f$, where $k_B T_f= \hbar^2 k_f^2/2$ and $k_f = (3 n \pi^2)^{1/3}$ are, respectively, the Fermi temperature and the Fermi wavevector of spin-unpolarized electron gas. The covered densities range between $r_s=2 \ldots 8 $ (presented in units of a$_0$) and finite temperatures between $\Theta=T/T_f=0.1 \ldots 1.0$.

\begin{figure}[t]
   \centering
   \includegraphics[width=8.6cm]{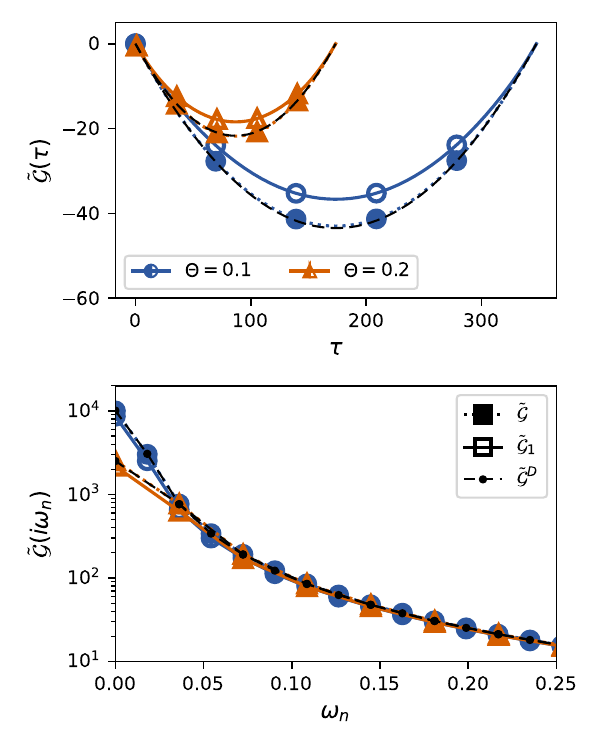}
   \caption{Comparison overview of the Drude model and the collective and 1-particle dipole correlation functions from PIMC, respectively, $\mathcal{\tilde G}$ and $\mathcal{\tilde G}_1$. In the top, the imaginary-time correlation functions $\mathcal{\tilde G}(\tau)$ and $\mathcal{\tilde G}_1$ are plotted at two temperatures, $\Theta_1=0.1$ and $\Theta_2=0.2$ and $r_s=8$ a$_0$, resulting in two curves ending at different values for $\beta_1 = 2 \beta_2$. In the bottom, the first few corresponding Matsubara data $\mathcal{\tilde G}(i \omega_n)$ are plotted, where the frequencies $\omega_n = 2 \pi n/\hbar \beta$ for $\Theta_1$ are twice those of $\Theta_2$, but the two functions lie on the same curve. In both cases, the values of $\mathcal{\tilde G}$ match the analytically continued Drude model within statistical uncertainties, whereas $\mathcal{\tilde G}_1$ is modestly weaker due to Coulomb compression.
   }
   \label{fig:analysis_overview}
\end{figure}

The main observables of interest are overviewed in Fig.~\ref{fig:analysis_overview}: the relative dipole autocorrelation function $\mathcal{\tilde G}(\tau)$ and its Matsubara series $\mathcal{\tilde G}(i \omega_n)$. As argued in Sec.~\ref{sec:theory}, the vertical offset of $\mathcal{\tilde G}(\tau)$ is arbitrary, and thus, we treat it as zero, making $\mathcal{\tilde G}(\tau) \leq 0$ for all $\tau$. Thus, data for the static Matsubara frequency would be numerically negative unless changed defined with an opposite sign like in Eq.~\eqref{eq:static_matsubara}. In the absence of notable many-body effects, all values of $\mathcal{\tilde G}(i \omega_n)$ lie on the same curve, except $\mathcal{\tilde G}(i \omega_n=0)$ which informs the first-order scale of the correlation function.

\begin{figure}[t]
   \centering
   \includegraphics[width=8.6cm]{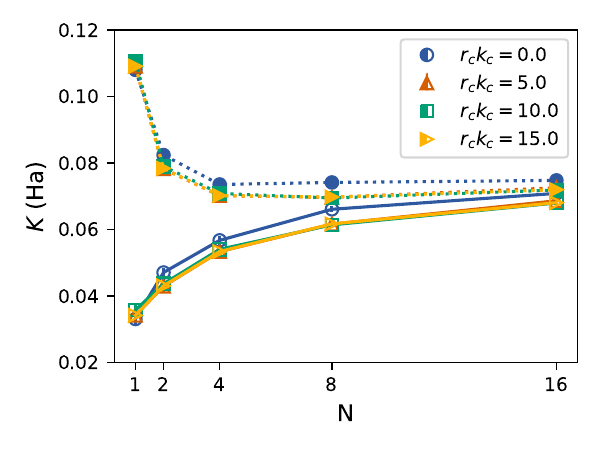}
   \\
   \includegraphics[width=8.6cm]{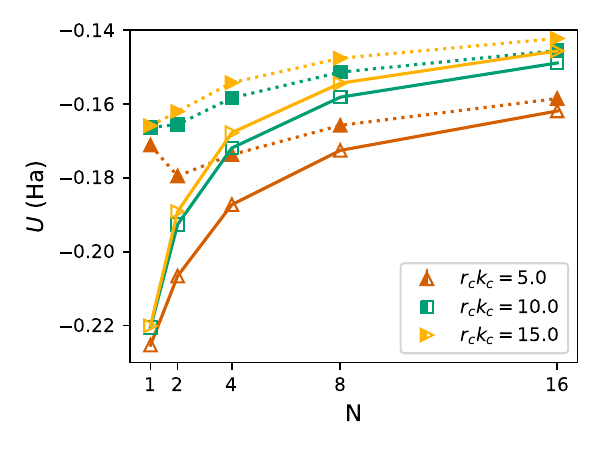}
   \caption{Scaling of per-particle kinetic energy $K$ (left) and potential energy $U$ (bottom) with finite $N$ and variable dimensional Ewald cutoff parameter $r_c k_c$. The solid lines with filled markers correspond to raw PIMC observables and the dotted lines with hollow markers are finite-size corrected values based on Eqs.~\eqref{eq:K_corr} and \eqref{eq:V_corr}. The case $r_c k_c=0$ means no analytical Ewald summation of any kind, including neutralizing background.
   }
   \label{fig:energy_vs_N}
\end{figure}

We generally seek traces of methodological and physical discrepancies between the PIMC data and the Drude response $\mathcal{\tilde G}^D$ due to the Coulomb correlation. The differences are treated in terms of relative error
\begin{equation}
    \label{eq:delta}
    \Delta \mathcal{\tilde G} = (\mathcal{\tilde G} - \mathcal{\tilde G}^D) / \mathcal{\tilde G}^D
\end{equation}
for any imaginary arguments $\tau$ or $i\omega_n$. The relative error is given in percent units. As the sign cancels out, $\Delta G$ is essentially a relative measure of the magnitude. Unless stated otherwise, we compare against the analytic expressions \eqref{eq:analytical_GiwD} and \eqref{eq:analytical_GtauD} in the $\Gamma \rightarrow 0$ limit.

On the one hand, the collective response $\mathcal{\tilde G}$ should always match $\mathcal{\tilde G}^D$ by virtue of perfect screening in long wavelength limit. Therefore, it conceals all subtle many-body phenomena of the simulation. Any statistically nonzero values of $\Delta \mathcal{\tilde G}$ or $\eta$, if they should be found, would only signal serious biases in methodology. On the other hand, the one-particle response $\mathcal{\tilde G}_1$ is a nontrivial many-body observable, and thus, a potent gauge of the subtleties of the PIMC simulation, such as biases due to finite time-step, finite size and finite Ewald summation.

\begin{figure*}[ht!]
   \centering
   \includegraphics[width=8.0cm]{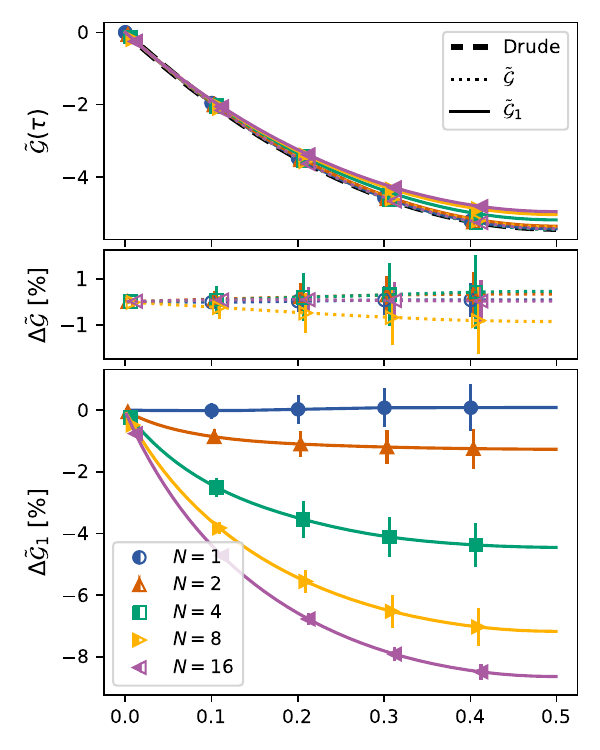}
   \includegraphics[width=8.0cm]{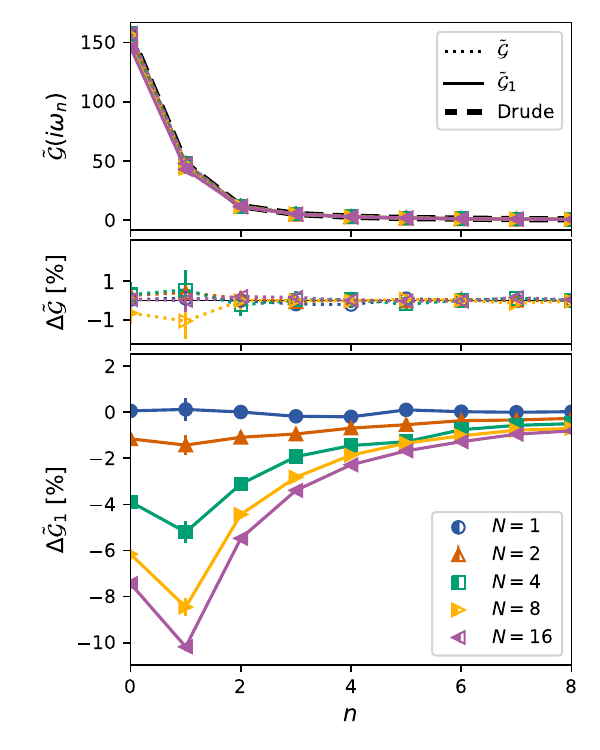}
   \caption{
   Scaling of the dipole autocorrelation function (left) and its Matsubara spectrum (right) with the number of particles $N$ in the PIMC simulation with $r_s=4.0$ and $\Theta=0.2$. The top panels of each figure show the absolute values of the collective responses $\mathcal{\tilde G}$ (dotted line, hollow markers), the 1-particle responses $\mathcal{\tilde G}_1$ (solid line, filled markers), and the Drude model (dashed line). Only selected markers and uncertainties of $\mathcal{\tilde G}(\tau)$ are plotted for clarity. The middle and lower panels show the relative differences of collective and one-particle estimators, respectively, based on Eq.~\eqref{eq:delta}.  The deviation $\Delta \mathcal{\tilde G}$ is persistently zero within uncertainties, whereas $\Delta \mathcal{\tilde G}_1$ grows monotonously toward the TDL with $N > 1$. For $N=1$, $\Delta \mathcal{\tilde G}_1$ remains zero because the simulation is evidently noninteracting.
   }
   \label{fig:mu2_vs_N}
\end{figure*}

However, let us first consider systematic merits of accuracy for the PIMC simulation based on energies. Finite-size effects of the energies are studied by varying the number of particles $N$ in the simulation, while keeping the density fixed. Results for kinetic and potential energies are presented in Fig.~\ref{fig:energy_vs_N} and for up to $N=16$ and $\Theta=0.2$ at $r_s=4$. Shown are also finite-size corrections \cite{Chiesa2006, Drummond2008, Brown2013} ($\xi=0$), which rectify the energies toward the thermodynamic limit (TDL), according to:
\begin{align}
    K_{p} &= \frac{K_N}{N} + \frac{1}{N}\left(\frac{\omega_p}{4} + \frac{5.264}{4 \pi r_s^2 (2 N)^{1/3}} \right) \label{eq:K_corr} \\
    U_{p} &= \frac{U_N}{N} + \frac{\omega_p}{4 N},
    \label{eq:V_corr}
\end{align}
where the corrected energies $K_p$ and $U_p$ are further multiplied with $\mathrm{tanh}(\omega_p \beta)$. The correlation energy is getting saturated with $N$, which also influences the kinetic energy. Simulation with $N > 16$, or a reliable extrapolation to the TDL, would be needed to resolve the statistically significant finite-size bias of the energies. Also, clearly the Ewald summation with $r_ck_c > 10$ is required to eliminate statistically significant bias from treating periodic images. Details of the Ewald parameters and a table of the energy values is found in the SI. 

The finite-size effects of the dipole polarizability are more topical: The scaling of $\mathcal{\tilde G}$ and $\mathcal{\tilde G}_1$ versus $N$ is plotted in Fig.~\ref{fig:mu2_vs_N} for $r_ck_c = 15.0$ at $r_s=4.0$ and $\Theta=0.2$. As expected, the finite-size effects for $\Delta \mathcal{\tilde G}$ are statistically zero within uncertainties of 1 \% or less, while the effects on $\Delta \mathcal{\tilde G}_1$ are immense. There is about 20~\% difference in the signal between $N=8$ and $N=16$, and so again, more than 16 particles would be required to saturate the finite-size effects to a reasonable level. Nevertheless, we will use $N=16$ for the remainder of this work. The effects due to finite time-step and the Ewald cutoff parameter $r_c k_c$ are also modest and only further discussed in the SI.

\begin{figure*}[t]
   \centering
   \includegraphics[width=8.0cm]{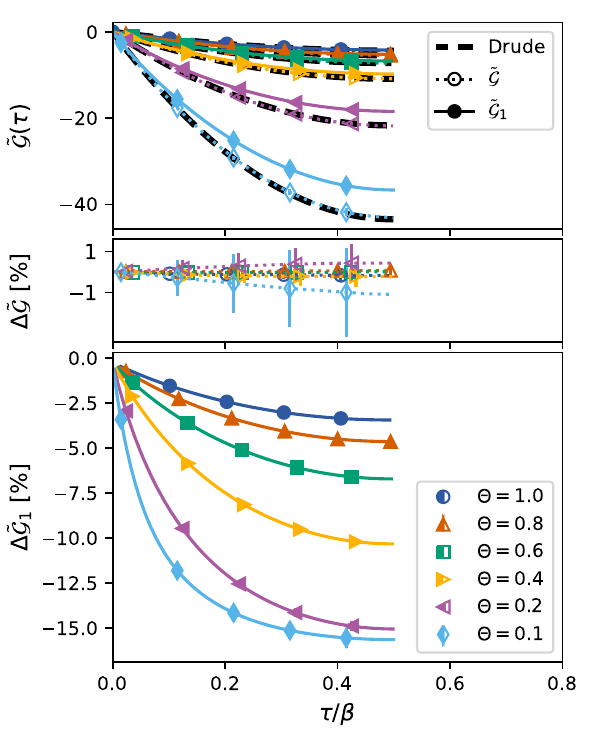}
   \includegraphics[width=8.0cm]{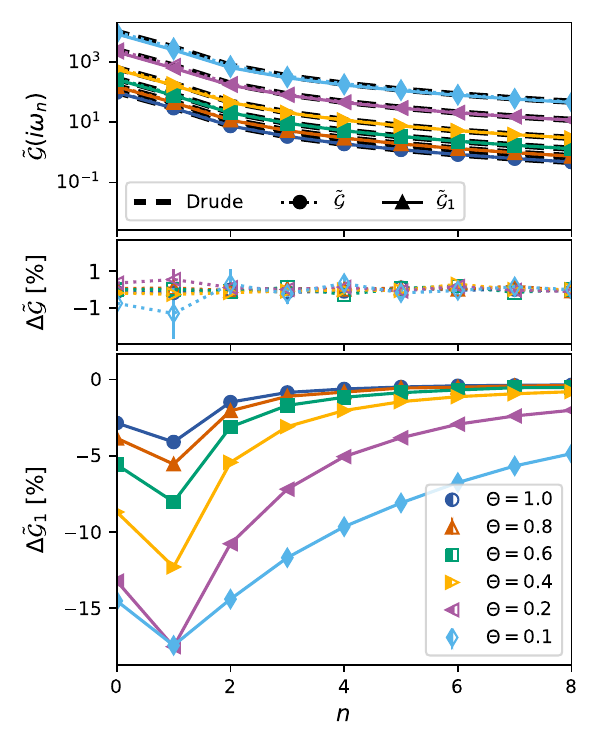}
   \caption{
   Scaling of the dipole autocorrelation function versus $\tau/\beta$ (left) and its Matsubara spectrum versus frequency index $n$ (right) is plotted with variable $\Theta$ in the PIMC simulation with $r_s=4.0$ and $N=16$. The top panels of each figure show the absolute values of the collective responses $\mathcal{\tilde G}$ (dotted line, hollow markers), the one-particle responses $\mathcal{\tilde G}_1$ (solid line, filled markers), and the Drude model (dashed line). The lower panels show the relative differences based on Eq.~\eqref{eq:delta}. Only selected values of $\mathcal{\tilde G}(\tau)$ are plotted for clarity. The deviation $\Delta \mathcal{\tilde G}$ is persistently zero within uncertainties, whereas $\Delta \mathcal{\tilde G}_1$ grows toward lower $\Theta$ with uncertainties smaller than the marker sizes.}
   \label{fig:Gtau_vs_theta}
\end{figure*}

\begin{figure}[h]
   \centering
   \includegraphics[width=8.6cm]{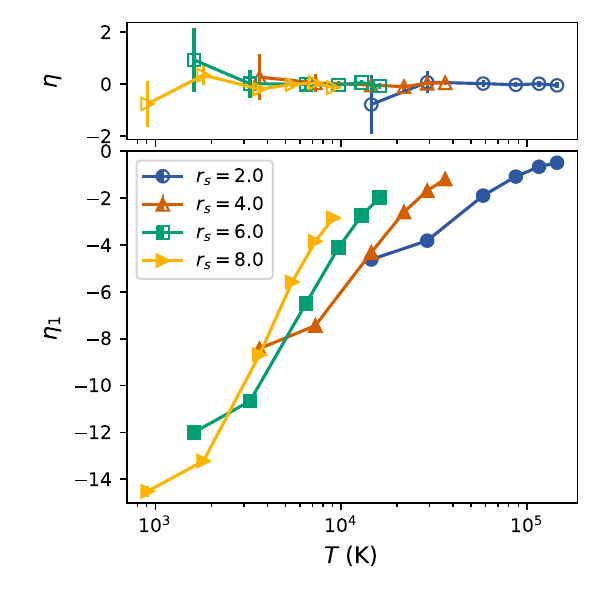}
   \caption{
   The match parameters for collective ($\eta$, upper panel; SI) and one-particle correlation functions ($\eta_1$, lower panel; Table~\ref{tab:eta_vs_rs_Theta}) are plotted versus temperature in different densities.}
   \label{fig:eta_vs_rs_theta}
\end{figure}

\begin{figure*}[h!]
   \centering
   \includegraphics[width=5.7cm]{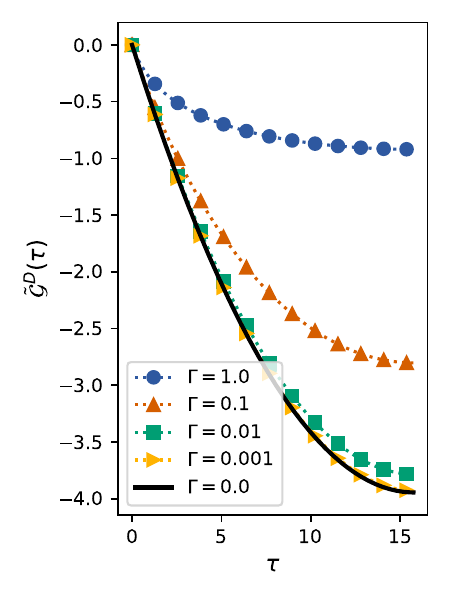}
   \includegraphics[width=5.7cm]{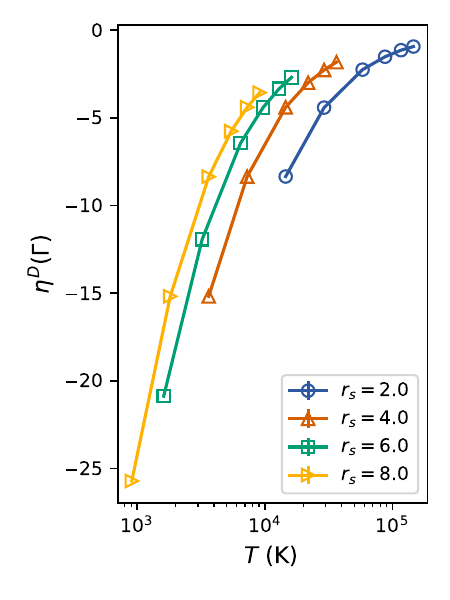}
   \includegraphics[width=5.7cm]{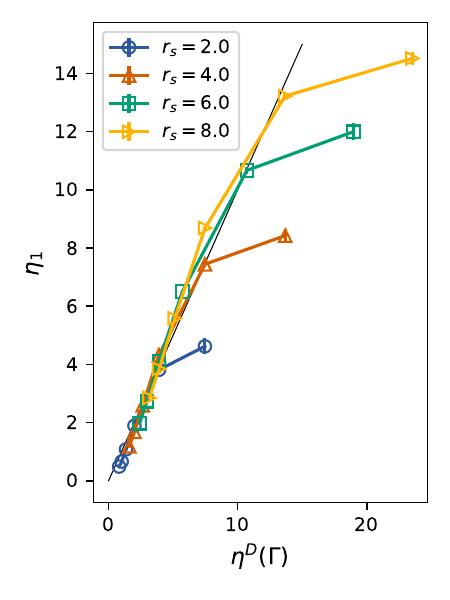}
   \caption{
   (left) Comparison of the $\mathcal{\tilde G}^D(\tau)$ with finite values for $\Gamma$. (middle) Values of $\eta^D(\Gamma = 0.065 / r_s)$ evaluated at the same temperatures and densities as the PIMC data $\eta_1$ in Fig.~\ref{fig:eta_vs_rs_theta}. (right) Comparison of the aforementioned $\eta^D$ vs $\eta_1$ with a solid black line indicating perfect match.
   }
   \label{fig:eta_vs_Gamma}
\end{figure*}

In Fig.~\ref{fig:Gtau_vs_theta} we plot $\mathcal{\tilde G}$ and $\mathcal{\tilde G}_1$ and their relative differences to the Drude model $\Delta \mathcal{\tilde G}$ versus $\Theta$ with $r_s=8$ and $N=16$. The plotting axes of $\mathcal{\tilde G}(\tau)$ are rescaled to $\tau/\beta$, emphasizing that the absolute values of $\mathcal{\tilde G}(\tau=\beta/2)$ decrease with the inverse temperature. Consequently, the Matsubara integral $\mathcal{\tilde G}(i\omega_n=0)$ goes down with $\beta$, but otherwise the Matsubara series $\mathcal{\tilde G}(i\omega_n > 0)$ follow the same curve, only with different spacings. Clearly, the collective response matches the reference in all temperatures as the relative difference $\Delta \mathcal{\tilde G}(\tau)$ stays zero within the uncertainties. In contrast, the one-particle response depends strongly on the temperature: the relative difference $\Delta \mathcal{\tilde G}_1(\tau)$ is magnified toward $\tau \rightarrow \beta / 2$, where it plateaus around $-16$ \% for the lowest temperature studied, $\Theta=0.1$. The same effect and approximate scaling carries over to the Matsubara spectrum $\Delta \mathcal{\tilde G}_1(i \omega_n)$. It should be noted that while $\Delta \mathcal{\tilde G}_1(i \omega_n)$ are plotted versus $n$, the data versus $\omega_n \propto \Theta$ follow approximately the same curve beyond $n > 0$. Nevertheless, the repulsive correlation embedded in $\Delta \mathcal{\tilde G}_1$ is bound to vanish in the high-frequency limit.

The qualitative picture conveyed by $\Delta \mathcal{\tilde G}_1 < 0$ is that the instantaneous dipole fluctuation of each particle is suppressed by the Coulomb pressure of its surroundings. The amount of the suppression scales with the ratio of potential versus kinetic energy, and thus, the effect starts to vanish when the temperature rises or when the density grows. For simplicity, let us monitor this suppression in the first order by
\begin{equation}
    \label{eq:eta}
    \eta \equiv \Delta \mathcal{\tilde G}(i \omega_n = 0),
\end{equation}
which measures the relative scaling in the numerical static limit. Values of $\eta$ are plotted in Fig.~\ref{fig:eta_vs_rs_theta} and presented in Table~\ref{tab:eta_vs_rs_Theta} for various densities and temperatures. Again, the plot summarizes that the collective response $\eta$ is in perfect agreement with the reference Drude model within uncertainties. On the one hand, this corroborates the perfect screening property of the ideal Coulomb plasma. On the other hand, this indicates absence of biases in the simulation procedure. Yet, it is the one-particle response $\eta_1$ that shows significant dependence on the physical parameters, density and temperature. The overall trend, for reasons argued above, is that the magnitude of $\eta_1$ grows toward lower temperatures and pressures.

So far we have made comparisons to the ideal Drude model with zero damping, $\Gamma \rightarrow 0$. However, we can identify effects on $\mathcal{\tilde G}^D$ with finite $\Gamma$ that appear similar to those with $\mathcal{\tilde G}_1$ immersed in the Coulomb pressure. This is illustrated in Fig.~\ref{fig:eta_vs_Gamma}. The first panel plots $\mathcal{\tilde G}^D(\tau)$ with various finite values for $\Gamma$. The plot shows that the finite phenomenological scattering rate has a similar suppressive effect on $\mathcal{\tilde G}_1(\tau)$ toward $\beta/2$. Therefore, one can make a phenomenological match between $\eta_1$ of PIMC and the Drude model $\eta^D(\Gamma > 0)$, where $\Gamma$ is the scattering rate from other electrons. Such a match can be approximated based on our data as 
\begin{equation}
    \label{eq:effective_Gamma1}
    \Gamma \approx 0.065 \tfrac{a_0}{r_s},
\end{equation}
which is plotted and compared to $\eta_1$ in Fig.~\ref{fig:eta_vs_Gamma}. For reference, observed scattering constants for ions are 1-2 orders of magnitude smaller: \textit{e.g.} $\Gamma_{\mathrm{Cu}} = 3.5 \times 10^{-3}$, $\Gamma_{\mathrm{Ag}} = 7.8 \times 10^{-4}$ and $\Gamma_{\mathrm{Au}} = 2.6 \times 10^{-3}$ \cite{Johnson1972} in atomic units. The simple model matches reasonably well to the high-temperature data, perhaps owing to the fact that the ratio of potential and kinetic energy scales approximately with $\sim r_s^{-1}$. The biggest discrepancies happen in low temperatures, where the PIMC correlation starts to plateau, which is not incorporated in Eq.~\eqref{eq:effective_Gamma1}. This could be attributed to the quantum zero-point motion, which typically arises in low-temperatures regimes of such simulations. However, the benchmark quality of our data in this regime should be taken with a grain of salt, due to low particle count and the lack Fermion exchange.

Overall, the consideration of qualitative and quantitative connections can be made more comprehensive between the phenomenological and \textit{ab initio} and the collective and one-particles pictures within each. Our presentation based on $\eta$ is a simple proof of concept that could be generalized, \textit{e.g.}, into a frequency spectrum to match the full spectra of $\mathcal{\tilde G}$ and not just their zeroth moments. Much like using static or dynamic LFCs of the momentum space, such information could be critical in mapping signals back from the imaginary to the real domain \cite{Groth2019}. We should stress, however, that our analysis in the long-wavelength limit does not straightforwardly connect with the Lindhard polarization picture expressed in Eq.~\eqref{eq:rpa}. However, because of the complicated nature of Eqs.~\eqref{eq:Aw_to_Gtau} and \eqref{eq:Aw_to_Giw}, it is possible that the best phenomenological model to express the one-particle response is something quite unlike the Drude model with Eq.~\eqref{eq:effective_Gamma1}, despite its superficial match to the PIMC data. Quests for the electrostatic picture and effective corrections shall be left for future.

\begin{table}[h]
    \centering
    \begin{tabular*}{\linewidth}{@{\extracolsep{\fill}} c|cccc}
            & \multicolumn{4}{c}{$r_s$}  \\
$\Theta$    &          2.0 &          4.0 &          6.0 &          8.0  \\ \hline
0.1         &       4.6(3) &       8.4(3) &      12.0(4) &      14.5(3)  \\
0.2         &     3.81(10) &     7.44(11) &    10.67(12) &    13.22(11)  \\
0.4         &      1.89(5) &      4.32(5) &      6.50(6) &      8.69(5)  \\
0.6         &      1.08(3) &      2.59(3) &      4.11(5) &      5.58(4)  \\
0.8         &      0.66(3) &      1.68(3) &      2.73(4) &      3.84(3)  \\
1.0         &      0.49(3) &      1.19(3) &      1.97(3) &      2.84(3)  \\
    \end{tabular*}
    \caption{The negative match parameter $-\eta_1$ (in percent units) for one-particle response with variable $r_s$ and $\Theta$ using $N=16$ and $r_ck_c=15$.}
    \label{tab:eta_vs_rs_Theta}
\end{table}

\section{Summary and outlook}
\label{sec:summary}

We have investigated PIMC estimators for the collective and one-particle dipole polarizabilities of periodic quantum plasmas in the long-wavelength limit, i.e., the optical regime. 
We have benchmarked the estimators with homogeneous Coulomb plasma in a range of finite densities ($r_s=2 \ldots 8$ a$_0$) and finite temperatures ($\Theta = 0.1 \ldots 1)$. 

The PIMC results are compared in the imaginary domain to the exact theoretical reference, namely the RPA or Lindhard response in $|q|\rightarrow 0$ limit, or the Drude model in the $\Gamma \rightarrow 0$ limit. The collective response matches the references statistically through ranges of physical and numerical parameters. This asserts the validity of the simulation in a weaker sense and provides yet another point of view to the perfect screening property of an ideal gas. This historical effect holds a steady collective appearance while emerging from numerous sensitive constituents, thus contrasting much of the emergent many-body physics of today.

In contrast, the one-particle responses show systematic dependencies on all parameters: the numerical ones (finite time-step, periodic images and number of particles) monitor accuracy of the many-body interactions, whereas the physical ones (finite density and temperature) unveil microscopic details of the optical response process. 

For a simple proof of concept, we trace this microscopic response to an effective Drude damping in a low-order approximation. However, the most convenient picture of local screening remains to be established. Clearly, it must be complementary to the LFC formalism of the RPA, whose many-body correction vanishes as $|q|\rightarrow 0$ \cite{kugler1975theory, Giuliani2008}. An educated formulation is also vital, if one wishes to face the challenges of analytical continuation to real domain, like LFC \cite{Groth2019, Hamann2020}. Therefore, it is highly convenient that this phenomenon can be self-consistently studied within the PIMC framework, even the same sampling. This opens speculation if one should seek this in the conventional structure factor formalism.

One might question what is the practical use of the optical response of individual particles, which cannot be directly measured. At least it should be recognized for its added insight and instrumental value in gauging many-body effects and self-consistent theories of local screening in the numerical simulation. Perhaps one could use it also to inform thermal density function theories \cite{Ramakrishna2020}. However, for fully accurate physical benchmark data one should seek Fermion statistics and mitigation of the finite-size effects with higher number of simulated particles and other corrective extrapolations.

Still, the relevant optical response that can be measured is the collective one, which however conceals all quantum many-body interactions by virtue of the perfect screening. Thus, we believe that having now firmly established the theoretical framework, numerical implementation, postprocessing, analyses and good practices (see Sec.~\ref{sec:pbc}), many future avenues open up:

i) Inversion to real domain. Where applicable, migration of numerical many-body correction from the integral transformations such as Eqs. \eqref{eq:Aw_to_Gtau} or \eqref{eq:Aw_to_Giw} is a formidable challenge. The maximum entropy approaches \cite{Gallicchio1994, Jarrell1996, Chuna2025} have been developed, but they work better for sharp Lorentzian peaks than the broad Drude-like spectrum \cite{Ceperley1995, Martin2016}. Finding analytical models and special approaches remains an open challenge.

ii) Confined spectrum. Treating interacting particles in environments less regular than a uniform potential (e.g. weak confinements, optical nanocavities or ionic plasma) should introduce non-trivial thermal and quantum effects for example with respect to the Drude--Lorentz model, characterized with resonance frequencies.

iii) Adsorption on surfaces. Dispersion interactions (\textit{e.g.} van der Waals and Casimir forces) between atoms and surfaces can be expressed with the instantaneous multipole fluctuations, \textit{i.e.}, polarizabilities in the Matsubara frequency domain \cite{Dalvit2011}. As the PIMC method treats them fluently \cite{Tiihonen2018, Tiihonen2019}, we may now also use it to inform effective adsorption models for Drude-like matter, such as HEG \cite{Tao2014}.

iv) Higher-order response properties. The response functions beyond the dipole order could be more sensitive to thermal quantum many-body effects. Their static and dynamic variants, such as $\chi^{(3)}$ of the Drude-like materials, could inform effective models \cite{Maack2018} and applications, such as ENZ materials \cite{Reshef2019}.

\section{Supplemental information}

The Supplemental information contains further benchmark figures of accuracy.

\begin{acknowledgments}
\label{sec:acknowledgements}
The authors wish to acknowledge CSC – IT Center for Science, Finland, and the Tampere Center for Scientific Computing; for computational resources. The authors acknowledge the financial support from the Photonics Research and Innovation Flagship (PREIN - decision 320165) and the Research Council of Finland project AQUA-PHOT (decision Grant No. 349350). JT acknowledges Ilkka Kylänpää for his support in the implementation and assessment of periodic potentials.
\end{acknowledgments}

\appendix
\section{Reduction of Lindhard theory to Drude model in the long-wavelength limit}
\label{sec:appendixA}

The Lindhard function in 3 dimensions, first mentioned in Sec. \ref{sec:intro}, is given by \cite{Giuliani2008}:
\begin{align}
    \chi_0(\textbf{q}, \omega) = \frac{1}{L^3} \sum \frac{f_{\textbf{k}+\textbf{q}}{-f_{\textbf{k}}}}{ E_{\textbf{k} + \textbf{q}} - E_{\textbf{k}} - \hbar(\omega + i\delta ) },
\end{align}
where $\chi_0(\textbf{q}, \omega)$ is the linear response to an incoming electric field with momentum $\textbf{q}$ and frequency $\omega$, $\textbf{k}$ and $E = \hbar^2k^2/2m_e$ are, respectively, the allowed momenta and associated energies in the electron gas, $m_e$ is the electron mass, $f_{\mathbf{k}}$ is the Fermi-Dirac distribution, and $\delta$ is an infinitesimal term associated to adiabatic switching-on. In the case of $G = 0$, the response given by Eq.~\eqref{eq:chi_q} is given by
\begin{equation}
\label{eq:Lindhard_susc}
\begin{split}
    \chi(\textbf{q}, \omega) =& - v_q\chi_0(\textbf{q}, \omega) \\
    =& -\frac{v_q}{L^3} \sum \frac{f_{\textbf{k}+\textbf{q}}{-f_{\textbf{k}}}}{ E_{\textbf{k} + \textbf{q}} - E_{\textbf{k}} - \hbar(\omega + i\delta ) },
\end{split}
\end{equation}
where $v_q$ is the reciprocal Coulomb potential. It is well known that this susceptibility approaches the Drude model with $\Gamma = 0$ by taking the limits of $q\rightarrow 0$ (known as the long-wavelength limit\cite{ashcroft, Singwi1968}) and $\delta \rightarrow 0$  \cite{haug2009quantum}. The following procedure recovers the Drude model with finite $\Gamma$, as define by Eq.~\eqref{eq:varepsilon_D}, by considering a finite $\delta$.

The numerator of the sum in Eq. \ref{eq:Lindhard_susc}  can be approximated as
\begin{equation}
    f_{\textbf{k} + \textbf{q}} - f_\textbf{k} \simeq  \textbf{q} \cdot \nabla_{\textbf{k}} f_\textbf{k},
\end{equation}
while the denominator can be written as
\begin{equation}
\begin{split}
    E_{\textbf{k} + \textbf{q} } - E_{\textbf{k}} &= \frac{\hbar^2}{2m_e}(k^2 + 2\textbf{k} \cdot \textbf{q} + q^2 ) - \frac{\hbar^2k^2}{2m_e} - \hbar(\omega + i\delta) \\
   & \simeq   \frac{\hbar^2 \textbf{k} \cdot \textbf{q}}{m_e} - \hbar(\omega + i\delta).
\end{split}
\end{equation}

Substituting these approximations in Eq. \ref{eq:Lindhard_susc} leads to:
\begin{equation}
    \chi(\textbf{q}, \omega) \simeq - \frac{v_q}{L^3} \sum_{\textbf{k}} \frac{\textbf{q}\cdot \nabla_\textbf{k}f_\textbf{k}}{\frac{\hbar^2 \textbf{k}\cdot\textbf{q}}{m_e}-\hbar( \omega + i\delta).}
\end{equation} 

Taking the factor $-\hbar(\omega + i\delta)$ outside of the sum and approximating the denominator with its Taylor expansion results in
\begin{equation} 
\label{eq:LindhardToDrudeSteps}
\begin{split}
    \chi(\textbf{q}, \omega)\simeq& \frac{v_q}{L^3\hbar(\omega + i\delta)} \sum_{\textbf{k}} \textbf{q}\cdot \nabla_\textbf{k}f_\textbf{k} \left( 1+\frac{\hbar \textbf{k}\cdot\textbf{q}}{m_e (\omega + i\delta)} \right) \\
    \simeq&  \frac{v_q}{L^3\hbar(\omega + i\delta)} \sum_{\textbf{k}} \textbf{q}\cdot \nabla_\textbf{k}f_\textbf{k} \frac{\hbar \textbf{k}\cdot\textbf{q}}{m_e (\omega + i\delta)} \\
    =& - \frac{v_qq^2}{L^3m_e (\omega + i\delta)^2} \sum_{\textbf{k}} f_\textbf{k} \\
  =& - \frac{e^2}{\varepsilon_0 q^2L^3} \frac{q^2N}{m_e (\omega + i\delta)^2}.\\
\end{split}
\end{equation} 

In Eq. \ref{eq:LindhardToDrudeSteps}, the term proportional to $\sum_\textbf{k}\nabla_\textbf{k}f_\textbf{k}$ vanishes and the remaining term can be solved by partial integration \cite{haug2009quantum}. Then, the sum $ \sum_{\textbf{k}} f_\textbf{k} $ corresponds to the number of electrons $N$. Finally, the fact that $v_q = e^2/|\textbf{q}|^2\varepsilon_0$ is used in the last step.

At this point, the dependency on $q$ is dropped and the quantity $\omega^2_p = (n e^2/\varepsilon_0 m_e)$ is recognized as the plasma frequency with number density $n = N/L^3$. Using these definitions leads to
\begin{equation}
  \chi(\omega) =    - \frac{\omega_p^2}{(\omega + i\delta)^2}.
\end{equation}

Finally, by defining $\delta = \Gamma / 2$ and considering $\delta^2 \approx 0$ ,  Eq. \ref{eq:chi_D} is recovered
\begin{align}
    \chi^{D}(\omega) = -\frac{\omega_p^2}{\omega^2 + i\omega \Gamma}.
\end{align}
We point that the fact that both $\delta$ in the Lindhard function and $\Gamma$ in the Drude model are infinitesimally small is the reason why we observe the best match in our results for small $\Gamma$.

\bibliography{references}

\end{document}